\documentclass[final,5p,times,onecolumn,unknownkeysallowed]{elsarticle}

\usepackage{doi}
\usepackage{hyperref}
\hypersetup{
	colorlinks=true,        
	linkcolor=blue,         
	citecolor=cyan,         
}

\usepackage{graphicx}
\usepackage{amssymb}
\usepackage{amsmath}
\usepackage{amsthm}
\usepackage{lineno}
\usepackage{hyperref}
\usepackage{multirow}
\usepackage{bigints}

\usepackage{dcolumn}
\usepackage{bm}
\usepackage{color}
\usepackage{xcolor}


\usepackage{dcolumn}
\usepackage{bm}
\usepackage{color}

\hypersetup{colorlinks=true,linkcolor=cyan,citecolor=cyan,urlcolor=blue,bookmarks=true}\modulolinenumbers[200]

\biboptions{comma,sort&compress}

\journal{Journal Name}

\begin{document}

\begin{frontmatter}

\title{Like Black holes, Buchdahl stars cannot be extremalized}

\author[mainaddress1,mainaddress2,mainaddress3,mainaddress4]
{Sanjar Shaymatov\cortext[cor2]{Corresponding author}\corref{cor2}}
\ead{sanjar@astrin.uz}
\author[mainaddress5]
{Naresh Dadhich}
\ead{nkd@iucaa.in}

\address[mainaddress1]{Institute for Theoretical Physics and Cosmology, Zhejiang University of Technology, Hangzhou 310023, China}
\address[mainaddress2]{Central Asian University,  Milliy Bog Street 264, Tashkent 111221, Uzbekistan}
\address[mainaddress3]{Institute of Fundamental and Applied Research, National Research University TIIAME, Kori Niyoziy 39, Tashkent 100000, Uzbekistan}
\address[mainaddress4]{National University of Uzbekistan, Tashkent 100174, Uzbekistan}
\address[mainaddress5]{Inter University Centre for Astronomy \&
Astrophysics, Post Bag 4, Pune 411007, India}

\date{Received: date / Accepted: date}

\begin{abstract}
It was shown long back in  \cite{Dadhich97} that a non-extremal black hole cannot be converted into an extremal one by test particle adiabatic accretion. The Buchdahl star is the most compact object without horizon and  is defined by $\Phi(R) = 4/9$, while black hole by $\Phi(R) = 1/2$. Here $\Phi(R)$ is the gravitational potential experienced by a particle, radially falling for static and axially for the rotating object. In this short note  we examine the question of extremalization of the Buchdhal star and show that the same result holds good as for the black hole. That is, a non-extremal Buchdahl star cannot be extremalized by test particle accretion. Further since extremal limit for BS is $>1$, it could facilitate formation of extremal black holes by neutral and spinless accretion. That is perhaps the only way they could be formed.

\end{abstract}


\end{frontmatter}

\linenumbers

\section{Introduction \label{Sec:introduction}} 

It turns out that a black hole (BH) is characterized by the gravitational potential, $\Phi(R) = 1/2$, while the most compact object without horizon \footnote{However black hole is primarily characterized the the geometric property of timlike Killing vector turning null, defining the Killing horizon. Buchdahl star can have no such geometric characterization. In more simple and pedestrian sense, the two are characterized by the potential.}, the Buchdahl star (BS) by $\Phi(R)=4/9$ \cite{Dadhich22,Pani22}. For the static case, $\Phi(R)$ is  the gravitational potential felt by a radially falling particle while for the rotating case, it is the potential felt by particle falling along the axis of rotation. In the latter case, one has to square out the centrifugal effect due to the dragging of inertial frame.   Thus we write $\Phi(R)=-\left({M-Q^2/2R}\right)/{R}$ for charged and $\Phi(R) =  -{M}/\big(R\left({1+a^2/R^2}\right)\big)$ for  rotating object.

In the static case, the Schwarzschild and Reissner-Nordstr{\"o}m metrics describe respectively a neutral and charged object irrespective of it being black hole or not. This is not so for the axially symmetric case where the Kerr metric describes only a rotating black hole and not a non-BH rotating object. This is because rotating object would suffer flattening at the poles giving rise to multipole moments which cannot be sustained by the Kerr metric having the spherical horizon. For BH, all multipole moments get evaporated away due to null character of horizon, which cannot be the case for a rotating star. As a matter of fact we have no exact solution describing a non-BH rotating object in general relativity. We have thus to resort to the Kerr metric for description of BS as an  approximation. This is perhaps a reasonable and acceptable approximation because BS is almost as compact as BH as its radius is very close to BH horizon. That is why we cannot exactly write the Smarr mass formula for BS. This is the overriding assumption we have to work with. 

Since BS is almost as compact as BH, it may therefore be expected to share many of the black hole properties. In particular we have recently investigated the weak cosmic censorship conjecture \cite{Shaymatov23JCAP} for BS and found that the BH result is carried over to BS as well. That is, WCCC may be violated at the  linear order but it is always restored when the second order perturbations are included \cite{Sorce-Wald17,An18,Ge18,Ning19,Shaymatov19c,Yan-Li19,Shaymatov21a,
Shaymatov20a,Jiang20plb,Shaymatov21d,Shaymatov22JCAP}. 

In the similar vein we would like to examine the question of extremalization; i.e., could a non-extremal BS be converted into an extremal one? It was shown in \cite{Wald74b} that an extremal black hole cannot be over-extremalized and further it was also shown in Ref.~\cite{Dadhich97} that a non-extremal black hole cannot be extremalized. In this letter we would establish the same result for BS. Like BH, a non-extremal BS cannot be converted into an extremal one nor can an extremal one be over-extremalised by test particle adiabatic accretion process. This happens because accreting energy $\delta M$ is bounded at both the ends. The bounds arise by requiring one, that infalling particle reaches the star surface and second, it tends to favour extremalisation. As extremality approaches the two bounds coincide, and so the parameter window of accreting particles pinches off.

This however does not rule out non-adiabatic discontinuous accretion that could however lead to over-extremality but never to extremality; i.e., the extremality could not be attained but it could however be jumped over. That is, there is a discontinuous jump. It is this phenomenon that gave rise to spate of intense activity in past some years on violation of weak cosmic censorship conjecture (WCCC)  \cite{Hubeny99,Matsas07PRL,Jacobson09,Saa11,Shaymatov15,Bouhmadi-Lopez10,Rocha14,Jana18,Song18,Duztas18,Duztas-Jamil20,Yang20a,Shaymatov19a,Gwak16JCAP,Gwak16PLB}. This all was however put to rest by Sorce and Wald \cite{Sorce-Wald17} by showing that when second order perturbations are included, the WCCC violation, that occurs at linear order accretion, is always restored  \cite{Sorce-Wald17,An18,Ge18,Ning19,Shaymatov19c,Yan-Li19,Shaymatov21a,Shaymatov20a,Jiang20plb,Shaymatov21d}, even for the Buchdahl star \cite{Shaymatov22JCAP}.

Since horizon blocks all information, there have been attempts to define some non-null surface which is close to horizon, for example the membrane paradigm \cite{Thorne86} and the stretched horizon \cite{Susskind93}. With this background, it is interesting that the Buchdahl star offers an excellent alternative as an astrophysical object which is almost as compact as BH without any apology or qualification, and could also share many of its properties. Its boundary is timelike and hence open to active physical interaction and accessible to external observer. There is therefore great merit and physical relevance in probing all the BH properties for BS. Here we would examine extremalization of charged and rotating BS and show that like BH it cannot be extremalized, and nor could extremal one be over-extremalized.

\section{Smarr mass formula \label{Sec:SS}}

For black hole we have the well known Smarr mass formula \cite{Smarr73,Bardeen73} which is given by
\begin{equation}
M = (\kappa/4\pi)A + 2\omega J + \Phi_e Q \, ,
\end{equation}
where $\kappa$, $\omega$ and $\Phi_e$ are respectively  surface gravity, frame dragging angular velocity and electromagnetic potential, all being evaluated at the horizon given by $R_{+} = M + \sqrt{M^2 - a^2 - Q^2}$, and $A$, $J$ ($a=J/M$) and $Q$ are the horizon area, angular momentum and electric charge respectively of BH. We would like to evaluate the above equation off the horizon at the Buchdahl radius $R_{BS} > R_{+}$. When they are evaluated off the horizon, they may not have the same thermodynamic meaning as they have for BH when evaluated at the horizon. \\

We begin by defining the gravitational potential felt in general by axially falling particle for charged and rotating object. This is to filter out the frame dragging effect due to rotation. The gravitational potential is then given by
\begin{equation}
\Phi(R) = \frac{M - Q^2/2R}{R\left(1+a^2/R^2\right)}\, .
\end{equation}
Its derivative would give surface gravity $\kappa$, which is the red-shifted proper acceleration relative to an asymptotic observer \cite{Thorne86} and is given by
\begin{equation}
\kappa = \frac{M_g(R)}{R^2+a^2}\, ,
\end{equation}
where
\begin{equation} \label{Eq:Mg}
M_g(R) = \frac{M\left(1-a^2/R^2\right)-Q^2/R}{1+a^2/R^2}.
\end{equation}
All this is to be evaluated at BH Horizon defined by $\Phi(R)=1/2$ and at the Buchdahl star radius given by $\Phi(R)=4/9$ \cite{Dadhich22, Pani22}. The former is given by
\begin{eqnarray}
R_{+} = M\left(1 + \sqrt{1-\alpha^2-\beta^2}\right)\, ,
\end{eqnarray}
while the latter by
\begin{eqnarray}
R_{BS} = \frac{9M}{8}\left(1 + \sqrt{1-\lambda^2}\right)
\end{eqnarray}
where $\lambda^2=(8/9)\alpha^2+(8/9)^2\beta^2$ and $\alpha^2=Q^2/M^2$, $\beta^2=a^2/M^2$ . \\

Now $M_g(R)$ and $\kappa$ are to be evaluated at $R_{+}$ and $R_{BS}$ for BH and BS. In particular $M_g(R_{+})= M\sqrt{1-\alpha^2-\beta^2}$ for BH while it is $M_g(R_{BS})= M\sqrt{1-\lambda^2}$ for BS. For BH, $\kappa =M_g(R_{+})/(R_{+}^2 + a^2)$, $A = 4\pi(R_{+}^2 + a^2)$,\, $\omega = a/(R_{+}^2 + a^2)$ and $\Phi_e = Q/R_{+}$, it is easy to see that all this verifies the above Smarr mass formula. \\

For BS, all this have to be evaluated at $R_{BS}>R_{+}$ and for which Kerr metric is only approximately applicable. It is therefore expected that the Smarr formula would not be exactly verified for non-BH rotating object. We begin by computing area and angular velocity,
\begin{eqnarray}\label{Eq.sur.ar.}
A&=&\int_{\Xi_{2}}\sqrt{det|g_{\alpha\beta}|}\,d\theta d\phi\nonumber\\&=&
\frac{2\pi M^2}{(M/R)^2}\int_{0}^{\pi}\bigg[\left(1+(M/R)^2\beta^2\right)^2\nonumber\\&-&\bigg(1-2(M/R)+(M/R)^2\beta^2\bigg)(M/R)^2\beta^2\sin^2\theta\bigg]^{1/2}\nonumber\\&\times & \sin\theta\, d\theta\, ,\\   
\label{first}
 \omega &=&\frac{2\,aMR}{R^2(R^2+a^2)+2MR\,a^2}|_{R_{BS}}\, ,
\end{eqnarray}
From the above equation the surface area yields
\begin{eqnarray}
A&=&
\frac{4\sqrt{2}\,\pi M^2}{64}\left(9+\gamma\right)^2\,\nonumber\\&\times & \int_{0}^{\pi}\sqrt{\frac{81 \big(81+9\gamma-32 \beta ^2\big)-16 (9+\gamma ) \beta ^2 \sin ^2\theta}{(9+\gamma)^4}}\nonumber\\&\times &\sin\theta\,d\theta\, ,
\end{eqnarray}
where we have defined $\gamma=\sqrt{81-64\beta^2}$.\\

For the static charged case, however the Smarr formula holds good at any R,
\begin{equation}
M = \frac{\kappa}{4\pi}A + \Phi_e Q = M - Q^2/R + Q^2/R\, .
\end{equation}

\begin{figure}
\centering
\includegraphics[scale=0.6]{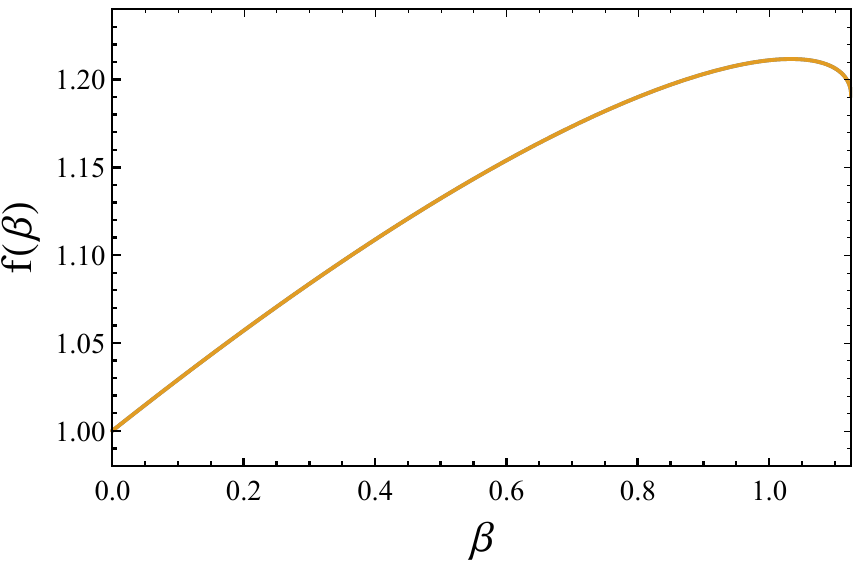}
\caption{\label{fig}
$f(\beta)$ is plotted against $\beta$.}
\end{figure}

On the other hand for the rotating BS, we write
\begin{eqnarray}\label{Eq:beta22}
f(\beta)&=&\frac{\kappa}{4\pi}A+2\left(\omega+\delta^{\prime}\right) J
\end{eqnarray}
where $\kappa$ is the surface gravity and  $\delta^{\prime}$ as given in Eq (13) below, when $\Delta \neq 0$ off the horizon. They are given by
\begin{eqnarray}
\kappa=\frac{(M/R)^2\left(1-(M/R)^2\beta^2\right)}{M\left(1+(M/R)^2\beta^2\right)^2}|_{R_{BS}}\, ,
\end{eqnarray}
and
\begin{eqnarray}
\delta^{\prime} &=&\frac{R^2\left(R^2-2MR+a^2\right)^{1/2}}{R^2(R^2+a^2)+2MR\,a^2}|_{R_{BS}}
 \, .
\label{second}
\end{eqnarray}
By using Eqs.~(\ref{first}-\ref{second}) in Eq.~(\ref{Eq:beta22}) we evaluate the Smarr formula at $R_{BS}$ as
\begin{eqnarray}
&f(\beta)&=\frac{\sqrt{2}}{162} M \big(81+9 \gamma -64 \beta ^2\big)\,\nonumber\\&\times & \int_{0}^{\pi}\bigg[\sqrt{\frac{81 \big(81+9\gamma-32 \beta ^2\big)-16 (9+\gamma ) \beta ^2 \sin ^2\theta}{(9+\gamma)^4}}\bigg]\nonumber\\&\times &\sin\theta\,d\theta
\nonumber\\&+&\frac{8 M \beta \left(64 \beta-\frac{32 \sqrt{2} \beta ^2}{\sqrt{9+\gamma}}+9 \sqrt{2} \sqrt{9+\gamma} \right)}{\left(81 \left(9+\gamma\right)-32 \beta ^2\right)}\, .
\end{eqnarray}
Note that $f(0.1)=1.02M$ and  $f(9/8)=1.26852M$ which clearly shows that the formula does not quite hold good off the horizon. This shows as expected off the horizon Smarr formula can hold only approximately. We plot $f(\beta)$ in Fig.~\ref{fig}. This cearly shows that the Smarr formula holds only at the horizon and not at the Buchdahl radius $R_{BS}>R_{+}$. 

\section{Extremalization \label{Sec:ex}}

As we have stated earlier that energy $\delta M$ of accreting particle is bounded at both the ends. It is well known that the lower bound comes from the equation of motion ensuring that particle reaches BH horizon or BS surface. The upper bound would come from the consideration that the accreting particle should tend BH/BS towards extremality. That is $dM_g(R)/dr \leq 0$. We now take up the charged and rotating cases one by one.

\subsection{Charged BS}

For charged object we can write the Smarr mass formula in the usual notation,
\begin{equation}
M = {\kappa/4\pi}A + Q\Phi_e = M_g(R) + Q^2/R \, ,
\end{equation}
where $\kappa = M_g(R)/R^2\, , A = 4\pi R^2\, , \Phi_e = Q/R$. This is valid for any arbitrary $R$, however $M_g(R_+) = M \sqrt{ 1- \alpha^2}$ for BH while for BS $\alpha^2$ is to be repleced by $8\alpha^2/9$. Note that $M_g(R) = M - Q^2/R$ which also follows from the Komar integral \cite{Komar1959} for the Reissner-Nordstr{\"o}m metric. The simple intuitive way to understand this is as follows: gravitational potential at $R$ would be $-(M-Q^2/2R)/R$, this is because electric field energy lying exterior to $R$ is to be subtracted from mass $M$. Then derivative of this potential gives the  gravitational acceleration as $-(M-Q^2/R)/R^2$. \\

Buchdahl had found the compactness bound $M/R \leq 4/9$ for a fluid star under very general conditions. It turns out that the bound could in general be written as $\Phi(R) \leq 4/9$ \cite{Dadhich20:JCAP, Dadhich22},  and the equality defines the Buchdahl star. For charged BS we then obtain
\begin{equation}
M/R = \frac{8/9}{1+\gamma} \, \, , \gamma^2 = 1-{8/9}\alpha^2 \, .
\end{equation}

For test particle accretion, the lower bound on $\delta M$ comes from the equation of motion for a radially falling particle,
\begin{equation}
\delta M\geq \frac{Q}{R}\delta Q \, , =
\frac{M}{R}\,\alpha\, \delta Q \, .
\end{equation}

On the other hand the upper bound follows from $\delta M_g \leq 0$ which implies
\begin{equation}
\delta M \leq (8/9)\, \alpha\, \delta Q \, .
\end{equation}

\begin{figure}
\centering
\includegraphics[scale=0.6]{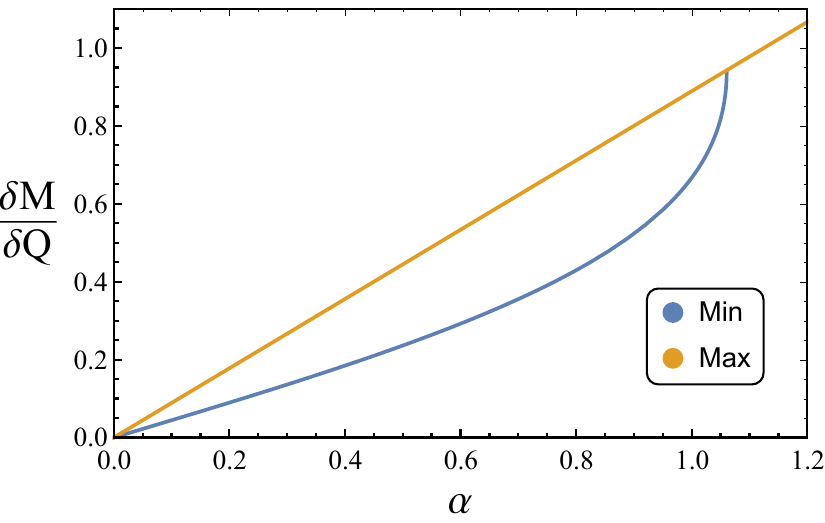}
\caption{\label{fig1} The bounds on  $\delta M/\delta Q$ against $\alpha$. }
\end{figure}

Taking the two together we write
\begin{equation}
\frac{(8/9)\alpha}{1+\gamma}\leq\frac{\delta M}{\delta Q}\leq(8/9)\,\alpha \ .
\end{equation}
As $\alpha^2 \to 9/8$; i.e., $\gamma \to 0$, both the bounds coincide and thereby implying that extremality can never be attained. This is because the parameter window for accreting particles pinches off as extremality is reached (See Fig.~\ref{fig1}). This is exactly what happens for the black hole \cite{Dadhich97}. It should however be noted that for Buchdahl star extremality bound is $\alpha^2 = 9/8 >1$ which is over-extremal for black hole. It is interesting that a non black hole object could have greater charge to mass ratio than black hole.

\subsection{Rotating BS}

We should bear in mind that the rotating case is more involved and has some important caveats. As stated earlier, the first and foremost is the fact that strictly speaking the Kerr metric describes only a rotating BH and not a non-BH rotating object. In the absence of anything better and the fact that Buchdahl rotating star would be very close to rotating black hole, we employ the Kerr metric for its description as a reasonable approximation. Secondly, we consider the equatorial plane test particle accretion for the maximum transmission of angular momentum to BS. Off the horizon computations become more involved and one has to resort to the numerical manipulation.\\

From the geodesic equation for test particle motion for the Kerr metric, we have
\begin{eqnarray}\label{Eq:motion}
E = \omega L + \sqrt{(\Delta/g_{\phi\phi})(L^2/g_{\phi\phi} +1)}\, ,
\end{eqnarray}
where $\Delta = R^2 -2MR + a^2$, $\,\omega = -g_{t\phi}/g_{\phi\phi}$. Here $E = \delta M$ and $L = \delta J$ are respectively energy and angular momentum of the accreting particle. Note that when $\Delta \neq 0$, the second term would also contribute to the first while considering the lower bound on $\delta M$ and hence we write
\begin{eqnarray}\label{Eq:inequality}
\delta M \geq (\omega + \delta^{\prime}) \delta J \,,
\end{eqnarray}
where $\omega$ and $\delta^{\prime}$ are given by Eqs.~(\ref{first}) and (\ref{second}).

\begin{figure}
\centering
\includegraphics[scale=0.6]{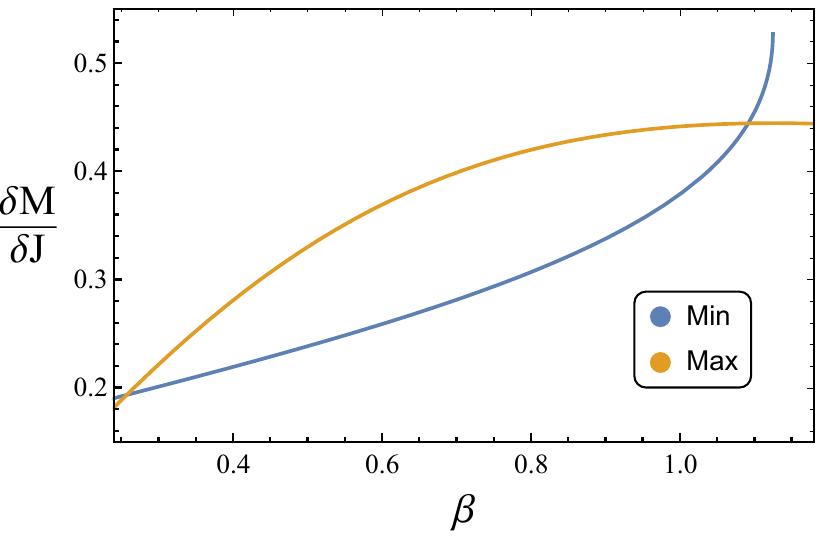}
\caption{\label{fig2}
The bounds on $\delta M/\delta J$  against $\beta$.
}
\end{figure}

Also we have from $\phi(R) = MR/(R^2 + a^2) = 4/9$,
\begin{eqnarray}\label{Eq:m/r}
M/R = \frac{8/9}{1+\sqrt{1-(8/9)^2\beta^{2}}}\, .
\end{eqnarray}
On substituting these into the above inequality (i.e., Eq.~(\ref{Eq:inequality})), we obtain the lower bound as
\begin{eqnarray}
&&\frac{\delta M}{\delta J} \geq \nonumber\\&&
\frac{4 \left(-\frac{32 \sqrt{2} \beta ^2}{\sqrt{9+\sqrt{81-64 \beta ^2}}}+9 \sqrt{2} \sqrt{\sqrt{81-64 \beta ^2}+9}+64 \beta \right)}{M\left(81 \left(9+\sqrt{81-64 \beta ^2}\right)-32 \beta ^2\right)}\, .\nonumber\\
 \end{eqnarray}
The upper bound follows from $dM_g/dr \leq 0$ where $M_g(R)$ is given in Eq.~(\ref{Eq:Mg}), and it would read as
\begin{eqnarray}
\frac{\delta M}{\delta J}\leq \frac{(8/9)^2\beta}{M\left(1+\left({8}/{9}\right)^2\beta^{2}\right)}\, .
\end{eqnarray}
Combining the two bounds we finally write
\begin{eqnarray}
&&
\frac{4 \left(-\frac{32 \sqrt{2} \beta ^2}{\sqrt{9+\sqrt{81-64 \beta ^2}}}+9 \sqrt{2} \sqrt{\sqrt{81-64 \beta ^2}+9}+64 \beta \right)}{M\Bigg(81 \left(9+\sqrt{81-64 \beta ^2}\right)-32 \beta ^2\Bigg)}\nonumber\\&&
 \leq \frac{\delta M}{\delta J}\leq \frac{(8/9)^2\beta}{M\left(1+\left({8}/{9}\right)^2\beta^{2}\right)}\, .
\end{eqnarray}
On numerical evaluation that we obtain for $\beta^{2} \to (9/8)^2$,
\begin{eqnarray}
\frac{0.5292}{M} \leq \frac{\delta M}{\delta J}\leq \frac{0.4444}{M}\, .
\end{eqnarray}
Here the lower bound rather than coinciding exceeds the upper one as $\beta^{2} \to  (9/8)^2$. This may be due to the approximations involved in evaluating $\omega$ and $\delta^{\prime}$. At any rate it bears out pinching off the parameter window for accreting particle and so rotating BS cannot be extremalized (See Fig.~\ref{fig2}).

\section{Over-extremalizing extremal BS \label{Sec:ov-ex}}
We now consider over-extremalization of extremal BS. We shall consider here the case of rotating BS and the same analysis could be carried through for the charged BS in a straightforward way. \\

Let us recall Eq.~(\ref{Eq:m/r})
\begin{eqnarray}
M/R = \frac{8/9}{1+\sqrt{1-\left(\frac{8}{9}\right)^2
\frac{a^2}{M^2}}}\, ,
\end{eqnarray}
For over-extremalization the minimum threshold of angular momentum would be
\begin{equation}
M^2 < \left(\frac{8}{9}\right)^2\,a^2\, .
\end{equation}
In the case of linear order accretion the above condition yields
\begin{equation}\label{naked00} \left({M+\delta M}\right)^{2}<
\left(\frac{8}{9}\right)^2\left(\frac{J+\delta J}{M+\delta M}\right)^2\, ,
\end{equation}
implying
\begin{equation}\label{naked} \left({M+\delta M}\right)^{2}<
\frac{8}{9}\left(J+\delta J\right) \, .
\end{equation}
We then obtain the lower bound of angular momentum required for over-extremalization of extremal rotating BS
\begin{eqnarray}
\delta J>\delta
J_{min}&=&\frac{9}{8}\Big[\left({M}^{2}-\frac{8}{9}J\right)+2 M
\delta M+\delta M^2\Big]\nonumber\\&=&
\frac{9}{4} M\,\delta M+\frac{9}{8}\delta M^2\, ,
\end{eqnarray}
where the extremality condition ${M}^{2}=({8}/{9})J$ is used.\\

Then the upper bound would be given by
\begin{equation}\label{Jmax} \delta J<\delta
J_{max}=\frac{R^2(R^2+a^2)+2MR\,a^2}{2\,aMR+R^2\left(R^2-2MR+a^2\right)^{1/2}}\vert_{R_{BS}}\, \delta M\, .
\end{equation}
On numerical evaluation we get
\begin{eqnarray}\label{jmax}
\delta J_{max}&=& \frac{153}{4\left(16+3\sqrt{2}\right)}M~\delta
M\, .
\end{eqnarray}
For over-extremaliation the difference, $\delta J = \delta J_{max} - \delta J_{min}$  is required to be positive. It reads as follows:
\begin{eqnarray}\label{minmax}
\delta J &=&\delta J_{max}-\delta J_{min} \nonumber\\&=& \frac{153}{4\left(16+3\sqrt{2}\right)}M~\delta
M-\frac{9}{4} M\,\delta M-\frac{9}{8}\delta M^2 \, .\nonumber\\ &=&1.89 M~\delta
M-\frac{9}{4} M\,\delta M -\frac{9}{8}\delta M^2 \, .
\end{eqnarray}
This clearly shows that $\delta J < 0$, and hence extremal rotating BS cannot be over-extremalized.

\section{Discussion\label{Sec:conclusion}}

The BS is the most compact astrophysical object without horizon. It is however fairly close to BH in compactness and perhaps in other properties as well. With this in mind we have recently also examined the validity of WCCC \cite{Shaymatov23JCAP}  and it turns out that the BH result is also carried forward to BS. That is, it could be violated at the linear order which is restored when second order perturbations are switched on. Continuing in the same vein here we have examined the question of extremalization of non-extremal and over-extremalization of of the extremal BS. Again the result turns out to be the same as for BH; i.e., neither non-extremal can be extremalized nor extremal over-extremalized. This is the main result.\\

For the static case, whatever holds for BH, which could, as it is, be taken over to BS, in particular the static vacuum solution describes both BH as well as any static object. That is the Smarr mass formula holds good at any arbitrary radius. However this is not true for the rotating case as we do not have a metric that describes a non-BH rotating object because the Kerr vacuum solution metric can only describe a BH and not a  rotating object. In the absence of the exact solution, we shall however use the Kerr metric for BS as an approximation. Similarly the other relevant geometric quantities arising from it like area $A$ and the frame-dragging angular velocity $\omega$ when evaluated off the horizon would suffer the same degree of approximation. That is why the Smarr formula does not exactly carries over to BS, however it could be taken as an approximation.\\

Since BS has extremal limit $>1$, it is overextremal relative to BH. This opens up an interesting possibility of forming an extremal BH by accretion. That is, BS could have charge/spin to mass ratio $>1$, by letting neutral or non-spinning particles impinge on and thereby reducing the ratio to extremal value $1$. That is, initially let 
$1<Q^2/M^2<9/8$ and then as neutral particles accrete, charge to mass ratio would reduce and may attain the extremal BH value, $Q^2/M^2=1$. On the other hand this is also possible that it continues to be BS with reduced charge/spin to mass ratio. It has recently been shown \cite{dago} that for that to happen it has to radiate out heat as the Vaidya radiation. That would be a specific tailored circumstance while  generic neutral/spinless accretion should give rise to an extremal BH. This is perhaps the only way extremal BH could be formed by accretion because non-extremal one cannot be extremalized. \\

Thus we reach another very important and remarkable conclusion that BS could serve as a precursor and faciltator to formation of extremal BHs by accretion. It is interesting that BH extremal limit cannot be reached from the bottom while it could be achieved from the top via the enabling agency of BS.  \\

On the other hand it is remarkable that the extremalization property studied here carries through wonderfully well for the rotating BS as well. It could in a straightforward manner be extended to the charged and rotating Kerr-Newman object.\\

There is however a basic difference between BH and BS in terms of their boundary, it is null for the former while timelike for the latter. That means BH simply swallows whatever that falls in without leaving any footprint in terms of scattering or reflection as nothing could emerge from the horizon. On the other hand timelike boundary of BS is two-way crossable and hence things could emerge out. All this would make the accretion process even less efficient for the extremalization as well as for the WCCC \cite{Shaymatov23JCAP}, and  thus it would work in favour of the results established in these two cases.\\

The BS is a naturally occurring real astrophysical object without any exotic stipulation and qualification. It therefore presents an excellent candidate as a BH mimicker, and it should thus be thoroughly probed for that. We believe it holds a great promise for exploring BH versus BS astrophysics, and see how do the two fare against the observations. The other most exciting aspect is the BH energetics. That is what we wish to take up next for rotating BS, in particular the Penrose process of energy extraction and its magnetic version -- magnetic Penrose process \cite{Wagh85,Dadhich18,Shaymatov22b} and superradiance in stars \cite{Richartz13,Cardoso15,Cardoso17,Day19JCAP,Chadha-Day22JCAP}.

\section*{Acknowledgments}
This work is supported by the National Natural Science Foundation of China under Grants No. 11675143 and No. 11975203, the National Key Research and Development Program of China under Grant No. 2020YFC2201503. ND wishes to thank Albert Einstein Institute, Golm  for the summer visit.

\bibliographystyle{apsrev4-1}  
\bibliography{gravreferences,ref}

\begin{thebibliography}{46}%
\makeatletter
\providecommand \@ifxundefined [1]{%
 \@ifx{#1\undefined}
}%
\providecommand \@ifnum [1]{%
 \ifnum #1\expandafter \@firstoftwo
 \else \expandafter \@secondoftwo
 \fi
}%
\providecommand \@ifx [1]{%
 \ifx #1\expandafter \@firstoftwo
 \else \expandafter \@secondoftwo
 \fi
}%
\providecommand \natexlab [1]{#1}%
\providecommand \enquote  [1]{``#1''}%
\providecommand \bibnamefont  [1]{#1}%
\providecommand \bibfnamefont [1]{#1}%
\providecommand \citenamefont [1]{#1}%
\providecommand \href@noop [0]{\@secondoftwo}%
\providecommand \href [0]{\begingroup \@sanitize@url \@href}%
\providecommand \@href[1]{\@@startlink{#1}\@@href}%
\providecommand \@@href[1]{\endgroup#1\@@endlink}%
\providecommand \@sanitize@url [0]{\catcode `\\12\catcode `\$12\catcode
  `\&12\catcode `\#12\catcode `\^12\catcode `\_12\catcode `\%12\relax}%
\providecommand \@@startlink[1]{}%
\providecommand \@@endlink[0]{}%
\providecommand \url  [0]{\begingroup\@sanitize@url \@url }%
\providecommand \@url [1]{\endgroup\@href {#1}{\urlprefix }}%
\providecommand \urlprefix  [0]{URL }%
\providecommand \Eprint [0]{\href }%
\providecommand \doibase [0]{http://dx.doi.org/}%
\providecommand \selectlanguage [0]{\@gobble}%
\providecommand \bibinfo  [0]{\@secondoftwo}%
\providecommand \bibfield  [0]{\@secondoftwo}%
\providecommand \translation [1]{[#1]}%
\providecommand \BibitemOpen [0]{}%
\providecommand \bibitemStop [0]{}%
\providecommand \bibitemNoStop [0]{.\EOS\space}%
\providecommand \EOS [0]{\spacefactor3000\relax}%
\providecommand \BibitemShut  [1]{\csname bibitem#1\endcsname}%
\let\auto@bib@innerbib\@empty
\bibitem [{\citenamefont {{Dadhich}}\ and\ \citenamefont
  {{Narayan}}(1997)}]{Dadhich97}%
  \BibitemOpen
  \bibfield  {author} {\bibinfo {author} {\bibfnamefont {N.}~\bibnamefont
  {{Dadhich}}}\ and\ \bibinfo {author} {\bibfnamefont {K.}~\bibnamefont
  {{Narayan}}},\ }\href {\doibase 10.1016/S0375-9601(97)00337-X} {\bibfield
  {journal} {\bibinfo  {journal} {Phys. Lett. A}\ }\textbf {\bibinfo {volume}
  {231}},\ \bibinfo {pages} {335} (\bibinfo {year} {1997})},\ \Eprint
  {http://arxiv.org/abs/gr-qc/9704070} {arXiv:gr-qc/9704070 [gr-qc]}
  \BibitemShut {NoStop}%
\bibitem [{\citenamefont {{Dadhich}}(2022)}]{Dadhich22}%
  \BibitemOpen
  \bibfield  {author} {\bibinfo {author} {\bibfnamefont {N.}~\bibnamefont
  {{Dadhich}}},\ }\href {\doibase 10.1103/PhysRevD.105.064044} {\bibfield
  {journal} {\bibinfo  {journal} {Phys. Rev. D}\ }\textbf {\bibinfo {volume}
  {105}},\ \bibinfo {eid} {064044} (\bibinfo {year} {2022})},\ \Eprint
  {http://arxiv.org/abs/2201.10381} {arXiv:2201.10381 [gr-qc]} \BibitemShut
  {NoStop}%
\bibitem [{\citenamefont {{Alho}}\ \emph {et~al.}(2022)\citenamefont {{Alho}},
  \citenamefont {{Nat{\'a}rio}}, \citenamefont {{Pani}},\ and\ \citenamefont
  {{Raposo}}}]{Pani22}%
  \BibitemOpen
  \bibfield  {author} {\bibinfo {author} {\bibfnamefont {A.}~\bibnamefont
  {{Alho}}}, \bibinfo {author} {\bibfnamefont {J.}~\bibnamefont
  {{Nat{\'a}rio}}}, \bibinfo {author} {\bibfnamefont {P.}~\bibnamefont
  {{Pani}}}, \ and\ \bibinfo {author} {\bibfnamefont {G.}~\bibnamefont
  {{Raposo}}},\ }\href@noop {} {\bibfield  {journal} {\bibinfo  {journal}
  {arXiv e-prints}\ ,\ \bibinfo {eid} {arXiv:2202.00043}} (\bibinfo {year}
  {2022})},\ \Eprint {http://arxiv.org/abs/2202.00043} {arXiv:2202.00043
  [gr-qc]} \BibitemShut {NoStop}%
\bibitem [{\citenamefont {{Shaymatov}}\ and\ \citenamefont
  {{Dadhich}}(2023)}]{Shaymatov23JCAP}%
  \BibitemOpen
  \bibfield  {author} {\bibinfo {author} {\bibfnamefont {S.}~\bibnamefont
  {{Shaymatov}}}\ and\ \bibinfo {author} {\bibfnamefont {N.}~\bibnamefont
  {{Dadhich}}},\ }\href {\doibase 10.1088/1475-7516/2023/06/010} {\bibfield
  {journal} {\bibinfo  {journal} {JCAP}\ }\textbf {\bibinfo {volume} {2023}},\
  \bibinfo {eid} {010} (\bibinfo {year} {2023})},\ \Eprint
  {http://arxiv.org/abs/2205.01350} {arXiv:2205.01350 [gr-qc]} \BibitemShut
  {NoStop}%
\bibitem [{\citenamefont {{Sorce}}\ and\ \citenamefont
  {{Wald}}(2017)}]{Sorce-Wald17}%
  \BibitemOpen
  \bibfield  {author} {\bibinfo {author} {\bibfnamefont {J.}~\bibnamefont
  {{Sorce}}}\ and\ \bibinfo {author} {\bibfnamefont {R.~M.}\ \bibnamefont
  {{Wald}}},\ }\href {\doibase 10.1103/PhysRevD.96.104014} {\bibfield
  {journal} {\bibinfo  {journal} {Phys. Rev. D}\ }\textbf {\bibinfo {volume}
  {96}},\ \bibinfo {eid} {104014} (\bibinfo {year} {2017})},\ \Eprint
  {http://arxiv.org/abs/1707.05862} {arXiv:1707.05862 [gr-qc]} \BibitemShut
  {NoStop}%
\bibitem [{\citenamefont {{An}}\ \emph {et~al.}(2018)\citenamefont {{An}},
  \citenamefont {{Shan}}, \citenamefont {{Zhang}},\ and\ \citenamefont
  {{Zhao}}}]{An18}%
  \BibitemOpen
  \bibfield  {author} {\bibinfo {author} {\bibfnamefont {J.}~\bibnamefont
  {{An}}}, \bibinfo {author} {\bibfnamefont {J.}~\bibnamefont {{Shan}}},
  \bibinfo {author} {\bibfnamefont {H.}~\bibnamefont {{Zhang}}}, \ and\
  \bibinfo {author} {\bibfnamefont {S.}~\bibnamefont {{Zhao}}},\ }\href
  {\doibase 10.1103/PhysRevD.97.104007} {\bibfield  {journal} {\bibinfo
  {journal} {Phys. Rev. D}\ }\textbf {\bibinfo {volume} {97}},\ \bibinfo {eid}
  {104007} (\bibinfo {year} {2018})},\ \Eprint
  {http://arxiv.org/abs/1711.04310} {arXiv:1711.04310 [hep-th]} \BibitemShut
  {NoStop}%
\bibitem [{\citenamefont {{Ge}}\ \emph {et~al.}(2018)\citenamefont {{Ge}},
  \citenamefont {{Mo}}, \citenamefont {{Zhao}},\ and\ \citenamefont
  {{Zheng}}}]{Ge18}%
  \BibitemOpen
  \bibfield  {author} {\bibinfo {author} {\bibfnamefont {B.}~\bibnamefont
  {{Ge}}}, \bibinfo {author} {\bibfnamefont {Y.}~\bibnamefont {{Mo}}}, \bibinfo
  {author} {\bibfnamefont {S.}~\bibnamefont {{Zhao}}}, \ and\ \bibinfo {author}
  {\bibfnamefont {J.}~\bibnamefont {{Zheng}}},\ }\href {\doibase
  10.1016/j.physletb.2018.07.015} {\bibfield  {journal} {\bibinfo  {journal}
  {Phys. Lett. B}\ }\textbf {\bibinfo {volume} {783}},\ \bibinfo {pages} {440}
  (\bibinfo {year} {2018})},\ \Eprint {http://arxiv.org/abs/1712.07342}
  {arXiv:1712.07342 [hep-th]} \BibitemShut {NoStop}%
\bibitem [{\citenamefont {{Ning}}\ \emph {et~al.}(2019)\citenamefont {{Ning}},
  \citenamefont {{Chen}},\ and\ \citenamefont {{Lin}}}]{Ning19}%
  \BibitemOpen
  \bibfield  {author} {\bibinfo {author} {\bibfnamefont {B.}~\bibnamefont
  {{Ning}}}, \bibinfo {author} {\bibfnamefont {B.}~\bibnamefont {{Chen}}}, \
  and\ \bibinfo {author} {\bibfnamefont {F.-L.}\ \bibnamefont {{Lin}}},\ }\href
  {\doibase 10.1103/PhysRevD.100.044043} {\bibfield  {journal} {\bibinfo
  {journal} {Phys. Rev. D}\ }\textbf {\bibinfo {volume} {100}},\ \bibinfo {eid}
  {044043} (\bibinfo {year} {2019})},\ \Eprint
  {http://arxiv.org/abs/1902.00949} {arXiv:1902.00949 [gr-qc]} \BibitemShut
  {NoStop}%
\bibitem [{\citenamefont {{Shaymatov}}\ \emph
  {et~al.}(2020{\natexlab{a}})\citenamefont {{Shaymatov}}, \citenamefont
  {{Dadhich}}, \citenamefont {{Ahmedov}},\ and\ \citenamefont
  {{Jamil}}}]{Shaymatov19c}%
  \BibitemOpen
  \bibfield  {author} {\bibinfo {author} {\bibfnamefont {S.}~\bibnamefont
  {{Shaymatov}}}, \bibinfo {author} {\bibfnamefont {N.}~\bibnamefont
  {{Dadhich}}}, \bibinfo {author} {\bibfnamefont {B.}~\bibnamefont
  {{Ahmedov}}}, \ and\ \bibinfo {author} {\bibfnamefont {M.}~\bibnamefont
  {{Jamil}}},\ }\href {\doibase 10.1140/epjc/s10052-020-8009-4} {\bibfield
  {journal} {\bibinfo  {journal} {Eur. Phys. J. C}\ }\textbf {\bibinfo {volume}
  {80}},\ \bibinfo {eid} {481} (\bibinfo {year} {2020}{\natexlab{a}})},\
  \Eprint {http://arxiv.org/abs/1908.01195} {arXiv:1908.01195 [gr-qc]}
  \BibitemShut {NoStop}%
\bibitem [{\citenamefont {{He}}\ and\ \citenamefont
  {{Jiang}}(2019)}]{Yan-Li19}%
  \BibitemOpen
  \bibfield  {author} {\bibinfo {author} {\bibfnamefont {Y.-L.}\ \bibnamefont
  {{He}}}\ and\ \bibinfo {author} {\bibfnamefont {J.}~\bibnamefont {{Jiang}}},\
  }\href {\doibase 10.1103/PhysRevD.100.124060} {\bibfield  {journal} {\bibinfo
   {journal} {Phys. Rev. D}\ }\textbf {\bibinfo {volume} {100}},\ \bibinfo
  {eid} {124060} (\bibinfo {year} {2019})},\ \Eprint
  {http://arxiv.org/abs/1912.05217} {arXiv:1912.05217 [hep-th]} \BibitemShut
  {NoStop}%
\bibitem [{\citenamefont {{Shaymatov}}\ and\ \citenamefont
  {{Dadhich}}(2021)}]{Shaymatov21a}%
  \BibitemOpen
  \bibfield  {author} {\bibinfo {author} {\bibfnamefont {S.}~\bibnamefont
  {{Shaymatov}}}\ and\ \bibinfo {author} {\bibfnamefont {N.}~\bibnamefont
  {{Dadhich}}},\ }\href {\doibase 10.1016/j.dark.2020.100758} {\bibfield
  {journal} {\bibinfo  {journal} {Phys. Dark Universe}\ }\textbf {\bibinfo
  {volume} {31}},\ \bibinfo {eid} {100758} (\bibinfo {year} {2021})},\ \Eprint
  {http://arxiv.org/abs/2004.09242} {arXiv:2004.09242 [gr-qc]} \BibitemShut
  {NoStop}%
\bibitem [{\citenamefont {{Shaymatov}}\ \emph
  {et~al.}(2020{\natexlab{b}})\citenamefont {{Shaymatov}}, \citenamefont
  {{Dadhich}},\ and\ \citenamefont {{Ahmedov}}}]{Shaymatov20a}%
  \BibitemOpen
  \bibfield  {author} {\bibinfo {author} {\bibfnamefont {S.}~\bibnamefont
  {{Shaymatov}}}, \bibinfo {author} {\bibfnamefont {N.}~\bibnamefont
  {{Dadhich}}}, \ and\ \bibinfo {author} {\bibfnamefont {B.}~\bibnamefont
  {{Ahmedov}}},\ }\href {\doibase 10.1103/PhysRevD.101.044028} {\bibfield
  {journal} {\bibinfo  {journal} {Phys. Rev. D}\ }\textbf {\bibinfo {volume}
  {101}},\ \bibinfo {eid} {044028} (\bibinfo {year} {2020}{\natexlab{b}})},\
  \Eprint {http://arxiv.org/abs/1908.07799} {arXiv:1908.07799 [gr-qc]}
  \BibitemShut {NoStop}%
\bibitem [{\citenamefont {{Jiang}}(2020)}]{Jiang20plb}%
  \BibitemOpen
  \bibfield  {author} {\bibinfo {author} {\bibfnamefont {J.}~\bibnamefont
  {{Jiang}}},\ }\href {\doibase 10.1016/j.physletb.2020.135365} {\bibfield
  {journal} {\bibinfo  {journal} {Phys. Lett. B}\ }\textbf {\bibinfo {volume}
  {804}},\ \bibinfo {eid} {135365} (\bibinfo {year} {2020})}\BibitemShut
  {NoStop}%
\bibitem [{\citenamefont {{Shaymatov}}\ \emph {et~al.}(2021)\citenamefont
  {{Shaymatov}}, \citenamefont {{Ahmedov}},\ and\ \citenamefont
  {{Jamil}}}]{Shaymatov21d}%
  \BibitemOpen
  \bibfield  {author} {\bibinfo {author} {\bibfnamefont {S.}~\bibnamefont
  {{Shaymatov}}}, \bibinfo {author} {\bibfnamefont {B.}~\bibnamefont
  {{Ahmedov}}}, \ and\ \bibinfo {author} {\bibfnamefont {M.}~\bibnamefont
  {{Jamil}}},\ }\href {\doibase 10.1140/epjc/s10052-021-09398-w} {\bibfield
  {journal} {\bibinfo  {journal} {Eur. Phys. J. C}\ }\textbf {\bibinfo {volume}
  {81}},\ \bibinfo {eid} {588} (\bibinfo {year} {2021})}\BibitemShut {NoStop}%
\bibitem [{\citenamefont {{Shaymatov}}\ and\ \citenamefont
  {{Dadhich}}(2022)}]{Shaymatov22JCAP}%
  \BibitemOpen
  \bibfield  {author} {\bibinfo {author} {\bibfnamefont {S.}~\bibnamefont
  {{Shaymatov}}}\ and\ \bibinfo {author} {\bibfnamefont {N.}~\bibnamefont
  {{Dadhich}}},\ }\href {\doibase 10.1088/1475-7516/2022/10/060} {\bibfield
  {journal} {\bibinfo  {journal} {JCAP}\ }\textbf {\bibinfo {volume} {2022}},\
  \bibinfo {eid} {060} (\bibinfo {year} {2022})},\ \Eprint
  {http://arxiv.org/abs/2008.04092} {arXiv:2008.04092 [gr-qc]} \BibitemShut
  {NoStop}%
\bibitem [{\citenamefont {{Wald}}(1974)}]{Wald74b}%
  \BibitemOpen
  \bibfield  {author} {\bibinfo {author} {\bibfnamefont {R.}~\bibnamefont
  {{Wald}}},\ }\href {\doibase 10.1016/0003-4916(74)90125-0} {\bibfield
  {journal} {\bibinfo  {journal} {Ann. Phys. (N.Y.)}\ }\textbf {\bibinfo
  {volume} {82}},\ \bibinfo {pages} {548} (\bibinfo {year} {1974})}\BibitemShut
  {NoStop}%
\bibitem [{\citenamefont {{Hubeny}}(1999)}]{Hubeny99}%
  \BibitemOpen
  \bibfield  {author} {\bibinfo {author} {\bibfnamefont {V.~E.}\ \bibnamefont
  {{Hubeny}}},\ }\href {\doibase 10.1103/PhysRevD.59.064013} {\bibfield
  {journal} {\bibinfo  {journal} {Phys. Rev. D}\ }\textbf {\bibinfo {volume}
  {59}},\ \bibinfo {eid} {064013} (\bibinfo {year} {1999})},\ \Eprint
  {http://arxiv.org/abs/gr-qc/9808043} {gr-qc/9808043} \BibitemShut {NoStop}%
\bibitem [{\citenamefont {{Matsas}}\ and\ \citenamefont {{da
  Silva}}(2007)}]{Matsas07PRL}%
  \BibitemOpen
  \bibfield  {author} {\bibinfo {author} {\bibfnamefont {G.~E.~A.}\
  \bibnamefont {{Matsas}}}\ and\ \bibinfo {author} {\bibfnamefont {A.~R.~R.}\
  \bibnamefont {{da Silva}}},\ }\href {\doibase 10.1103/PhysRevLett.99.181301}
  {\bibfield  {journal} {\bibinfo  {journal} {Phys. Rev. Lett.}\ }\textbf
  {\bibinfo {volume} {99}},\ \bibinfo {eid} {181301} (\bibinfo {year}
  {2007})},\ \Eprint {http://arxiv.org/abs/0706.3198} {arXiv:0706.3198 [gr-qc]}
  \BibitemShut {NoStop}%
\bibitem [{\citenamefont {{Jacobson}}\ and\ \citenamefont
  {{Sotiriou}}(2009)}]{Jacobson09}%
  \BibitemOpen
  \bibfield  {author} {\bibinfo {author} {\bibfnamefont {T.}~\bibnamefont
  {{Jacobson}}}\ and\ \bibinfo {author} {\bibfnamefont {T.~P.}\ \bibnamefont
  {{Sotiriou}}},\ }\href {\doibase 10.1103/PhysRevLett.103.141101} {\bibfield
  {journal} {\bibinfo  {journal} {Phys. Rev. Lett.}\ }\textbf {\bibinfo
  {volume} {103}},\ \bibinfo {eid} {141101} (\bibinfo {year} {2009})},\ \Eprint
  {http://arxiv.org/abs/0907.4146} {arXiv:0907.4146 [gr-qc]} \BibitemShut
  {NoStop}%
\bibitem [{\citenamefont {{Saa}}\ and\ \citenamefont
  {{Santarelli}}(2011)}]{Saa11}%
  \BibitemOpen
  \bibfield  {author} {\bibinfo {author} {\bibfnamefont {A.}~\bibnamefont
  {{Saa}}}\ and\ \bibinfo {author} {\bibfnamefont {R.}~\bibnamefont
  {{Santarelli}}},\ }\href {\doibase 10.1103/PhysRevD.84.027501} {\bibfield
  {journal} {\bibinfo  {journal} {Phys. Rev. D}\ }\textbf {\bibinfo {volume}
  {84}},\ \bibinfo {eid} {027501} (\bibinfo {year} {2011})},\ \Eprint
  {http://arxiv.org/abs/1105.3950} {arXiv:1105.3950 [gr-qc]} \BibitemShut
  {NoStop}%
\bibitem [{\citenamefont {{Shaymatov}}\ \emph {et~al.}(2015)\citenamefont
  {{Shaymatov}}, \citenamefont {{Patil}}, \citenamefont {{Ahmedov}},\ and\
  \citenamefont {{Joshi}}}]{Shaymatov15}%
  \BibitemOpen
  \bibfield  {author} {\bibinfo {author} {\bibfnamefont {S.}~\bibnamefont
  {{Shaymatov}}}, \bibinfo {author} {\bibfnamefont {M.}~\bibnamefont
  {{Patil}}}, \bibinfo {author} {\bibfnamefont {B.}~\bibnamefont {{Ahmedov}}},
  \ and\ \bibinfo {author} {\bibfnamefont {P.~S.}\ \bibnamefont {{Joshi}}},\
  }\href {\doibase 10.1103/PhysRevD.91.064025} {\bibfield  {journal} {\bibinfo
  {journal} {Phys. Rev. D}\ }\textbf {\bibinfo {volume} {91}},\ \bibinfo {eid}
  {064025} (\bibinfo {year} {2015})},\ \Eprint {http://arxiv.org/abs/1409.3018}
  {arXiv:1409.3018 [gr-qc]} \BibitemShut {NoStop}%
\bibitem [{\citenamefont {{Bouhmadi-L{\'o}pez}}\ \emph
  {et~al.}(2010)\citenamefont {{Bouhmadi-L{\'o}pez}}, \citenamefont
  {{Cardoso}}, \citenamefont {{Nerozzi}},\ and\ \citenamefont
  {{Rocha}}}]{Bouhmadi-Lopez10}%
  \BibitemOpen
  \bibfield  {author} {\bibinfo {author} {\bibfnamefont {M.}~\bibnamefont
  {{Bouhmadi-L{\'o}pez}}}, \bibinfo {author} {\bibfnamefont {V.}~\bibnamefont
  {{Cardoso}}}, \bibinfo {author} {\bibfnamefont {A.}~\bibnamefont
  {{Nerozzi}}}, \ and\ \bibinfo {author} {\bibfnamefont {J.~V.}\ \bibnamefont
  {{Rocha}}},\ }\href {\doibase 10.1103/PhysRevD.81.084051} {\bibfield
  {journal} {\bibinfo  {journal} {Phys. Rev. D}\ }\textbf {\bibinfo {volume}
  {81}},\ \bibinfo {eid} {084051} (\bibinfo {year} {2010})},\ \Eprint
  {http://arxiv.org/abs/1003.4295} {arXiv:1003.4295 [gr-qc]} \BibitemShut
  {NoStop}%
\bibitem [{\citenamefont {{Rocha}}\ and\ \citenamefont
  {{Santarelli}}(2014)}]{Rocha14}%
  \BibitemOpen
  \bibfield  {author} {\bibinfo {author} {\bibfnamefont {J.~V.}\ \bibnamefont
  {{Rocha}}}\ and\ \bibinfo {author} {\bibfnamefont {R.}~\bibnamefont
  {{Santarelli}}},\ }\href {\doibase 10.1103/PhysRevD.89.064065} {\bibfield
  {journal} {\bibinfo  {journal} {Phys. Rev. D}\ }\textbf {\bibinfo {volume}
  {89}},\ \bibinfo {eid} {064065} (\bibinfo {year} {2014})},\ \Eprint
  {http://arxiv.org/abs/1402.4840} {arXiv:1402.4840 [gr-qc]} \BibitemShut
  {NoStop}%
\bibitem [{\citenamefont {{Jana}}\ \emph {et~al.}(2018)\citenamefont {{Jana}},
  \citenamefont {{Shaikh}},\ and\ \citenamefont {{Sarkar}}}]{Jana18}%
  \BibitemOpen
  \bibfield  {author} {\bibinfo {author} {\bibfnamefont {S.}~\bibnamefont
  {{Jana}}}, \bibinfo {author} {\bibfnamefont {R.}~\bibnamefont {{Shaikh}}}, \
  and\ \bibinfo {author} {\bibfnamefont {S.}~\bibnamefont {{Sarkar}}},\ }\href
  {\doibase 10.1103/PhysRevD.98.124039} {\bibfield  {journal} {\bibinfo
  {journal} {Phys. Rev. D}\ }\textbf {\bibinfo {volume} {98}},\ \bibinfo {eid}
  {124039} (\bibinfo {year} {2018})},\ \Eprint
  {http://arxiv.org/abs/1808.09656} {arXiv:1808.09656 [gr-qc]} \BibitemShut
  {NoStop}%
\bibitem [{\citenamefont {{Song}}\ \emph {et~al.}(2018)\citenamefont {{Song}},
  \citenamefont {{Zhang}}, \citenamefont {{Zou}}, \citenamefont {{Sun}},\ and\
  \citenamefont {{Yue}}}]{Song18}%
  \BibitemOpen
  \bibfield  {author} {\bibinfo {author} {\bibfnamefont {Y.}~\bibnamefont
  {{Song}}}, \bibinfo {author} {\bibfnamefont {M.}~\bibnamefont {{Zhang}}},
  \bibinfo {author} {\bibfnamefont {D.-C.}\ \bibnamefont {{Zou}}}, \bibinfo
  {author} {\bibfnamefont {C.-Y.}\ \bibnamefont {{Sun}}}, \ and\ \bibinfo
  {author} {\bibfnamefont {R.-H.}\ \bibnamefont {{Yue}}},\ }\href {\doibase
  10.1088/0253-6102/69/6/694} {\bibfield  {journal} {\bibinfo  {journal}
  {Commun. Theor. Phys.}\ }\textbf {\bibinfo {volume} {69}},\ \bibinfo {pages}
  {694} (\bibinfo {year} {2018})},\ \Eprint {http://arxiv.org/abs/1705.01676}
  {arXiv:1705.01676 [gr-qc]} \BibitemShut {NoStop}%
\bibitem [{\citenamefont {{D{\"u}zta{\c s}}}(2018)}]{Duztas18}%
  \BibitemOpen
  \bibfield  {author} {\bibinfo {author} {\bibfnamefont {K.}~\bibnamefont
  {{D{\"u}zta{\c s}}}},\ }\href {\doibase 10.1088/1361-6382/aaa4e0} {\bibfield
  {journal} {\bibinfo  {journal} {Class. Quantum Grav.}\ }\textbf {\bibinfo
  {volume} {35}},\ \bibinfo {eid} {045008} (\bibinfo {year} {2018})},\ \Eprint
  {http://arxiv.org/abs/1710.06610} {arXiv:1710.06610 [gr-qc]} \BibitemShut
  {NoStop}%
\bibitem [{\citenamefont {{D{\"u}zta{\c{s}}}}\ \emph
  {et~al.}(2020)\citenamefont {{D{\"u}zta{\c{s}}}}, \citenamefont {{Jamil}},
  \citenamefont {{Shaymatov}},\ and\ \citenamefont
  {{Ahmedov}}}]{Duztas-Jamil20}%
  \BibitemOpen
  \bibfield  {author} {\bibinfo {author} {\bibfnamefont {K.}~\bibnamefont
  {{D{\"u}zta{\c{s}}}}}, \bibinfo {author} {\bibfnamefont {M.}~\bibnamefont
  {{Jamil}}}, \bibinfo {author} {\bibfnamefont {S.}~\bibnamefont
  {{Shaymatov}}}, \ and\ \bibinfo {author} {\bibfnamefont {B.}~\bibnamefont
  {{Ahmedov}}},\ }\href {\doibase 10.1088/1361-6382/ab9d96} {\bibfield
  {journal} {\bibinfo  {journal} {Class. Quantum Grav.}\ }\textbf {\bibinfo
  {volume} {37}},\ \bibinfo {pages} {175005} (\bibinfo {year} {2020})},\
  \Eprint {http://arxiv.org/abs/1808.04711} {arXiv:1808.04711 [gr-qc]}
  \BibitemShut {NoStop}%
\bibitem [{\citenamefont {{Yang}}\ \emph {et~al.}(2020)\citenamefont {{Yang}},
  \citenamefont {{Chen}}, \citenamefont {{Wan}}, \citenamefont {{Wei}},\ and\
  \citenamefont {{Liu}}}]{Yang20a}%
  \BibitemOpen
  \bibfield  {author} {\bibinfo {author} {\bibfnamefont {S.-J.}\ \bibnamefont
  {{Yang}}}, \bibinfo {author} {\bibfnamefont {J.}~\bibnamefont {{Chen}}},
  \bibinfo {author} {\bibfnamefont {J.-J.}\ \bibnamefont {{Wan}}}, \bibinfo
  {author} {\bibfnamefont {S.-W.}\ \bibnamefont {{Wei}}}, \ and\ \bibinfo
  {author} {\bibfnamefont {Y.-X.}\ \bibnamefont {{Liu}}},\ }\href {\doibase
  10.1103/PhysRevD.101.064048} {\bibfield  {journal} {\bibinfo  {journal}
  {Phys. Rev. D}\ }\textbf {\bibinfo {volume} {101}},\ \bibinfo {eid} {064048}
  (\bibinfo {year} {2020})},\ \Eprint {http://arxiv.org/abs/2001.03106}
  {arXiv:2001.03106 [gr-qc]} \BibitemShut {NoStop}%
\bibitem [{\citenamefont {{Shaymatov}}\ \emph {et~al.}(2019)\citenamefont
  {{Shaymatov}}, \citenamefont {{Dadhich}},\ and\ \citenamefont
  {{Ahmedov}}}]{Shaymatov19a}%
  \BibitemOpen
  \bibfield  {author} {\bibinfo {author} {\bibfnamefont {S.}~\bibnamefont
  {{Shaymatov}}}, \bibinfo {author} {\bibfnamefont {N.}~\bibnamefont
  {{Dadhich}}}, \ and\ \bibinfo {author} {\bibfnamefont {B.}~\bibnamefont
  {{Ahmedov}}},\ }\href {\doibase 10.1140/epjc/s10052-019-7088-6} {\bibfield
  {journal} {\bibinfo  {journal} {Eur. Phys. J. C}\ }\textbf {\bibinfo {volume}
  {79}},\ \bibinfo {pages} {585} (\bibinfo {year} {2019})},\ \Eprint
  {http://arxiv.org/abs/1809.10457} {arXiv:1809.10457 [gr-qc]} \BibitemShut
  {NoStop}%
\bibitem [{\citenamefont {{Gwak}}\ and\ \citenamefont
  {{Lee}}(2016{\natexlab{a}})}]{Gwak16JCAP}%
  \BibitemOpen
  \bibfield  {author} {\bibinfo {author} {\bibfnamefont {B.}~\bibnamefont
  {{Gwak}}}\ and\ \bibinfo {author} {\bibfnamefont {B.-H.}\ \bibnamefont
  {{Lee}}},\ }\href {\doibase 10.1088/1475-7516/2016/02/015} {\bibfield
  {journal} {\bibinfo  {journal} {JCAP}\ }\textbf {\bibinfo {volume} {2016}},\
  \bibinfo {eid} {015} (\bibinfo {year} {2016}{\natexlab{a}})},\ \Eprint
  {http://arxiv.org/abs/1509.06691} {arXiv:1509.06691 [gr-qc]} \BibitemShut
  {NoStop}%
\bibitem [{\citenamefont {{Gwak}}\ and\ \citenamefont
  {{Lee}}(2016{\natexlab{b}})}]{Gwak16PLB}%
  \BibitemOpen
  \bibfield  {author} {\bibinfo {author} {\bibfnamefont {B.}~\bibnamefont
  {{Gwak}}}\ and\ \bibinfo {author} {\bibfnamefont {B.-H.}\ \bibnamefont
  {{Lee}}},\ }\href {\doibase 10.1016/j.physletb.2016.02.028} {\bibfield
  {journal} {\bibinfo  {journal} {Phys. Let. B}\ }\textbf {\bibinfo {volume}
  {755}},\ \bibinfo {pages} {324} (\bibinfo {year} {2016}{\natexlab{b}})},\
  \Eprint {http://arxiv.org/abs/1510.08215} {arXiv:1510.08215 [gr-qc]}
  \BibitemShut {NoStop}%
\bibitem [{\citenamefont {Thorne}\ \emph {et~al.}(1986)\citenamefont {Thorne},
  \citenamefont {Price},\ and\ \citenamefont {Macdonald}}]{Thorne86}%
  \BibitemOpen
  \bibinfo {editor} {\bibfnamefont {K.~S.}\ \bibnamefont {Thorne}}, \bibinfo
  {editor} {\bibfnamefont {R.~H.}\ \bibnamefont {Price}}, \ and\ \bibinfo
  {editor} {\bibfnamefont {D.~A.}\ \bibnamefont {Macdonald}},\ eds.,\
  \href@noop {} {\emph {\bibinfo {title} {Black Holes: The Membrane
  Paradigm}}}\ (\bibinfo  {publisher} {Yale University Press},\ \bibinfo
  {address} {London},\ \bibinfo {year} {1986})\BibitemShut {NoStop}%
\bibitem [{\citenamefont {{Susskind}}\ \emph {et~al.}(1993)\citenamefont
  {{Susskind}}, \citenamefont {{Thorlacius}},\ and\ \citenamefont
  {{Uglum}}}]{Susskind93}%
  \BibitemOpen
  \bibfield  {author} {\bibinfo {author} {\bibfnamefont {L.}~\bibnamefont
  {{Susskind}}}, \bibinfo {author} {\bibfnamefont {L.}~\bibnamefont
  {{Thorlacius}}}, \ and\ \bibinfo {author} {\bibfnamefont {J.}~\bibnamefont
  {{Uglum}}},\ }\href {\doibase 10.1103/PhysRevD.48.3743} {\bibfield  {journal}
  {\bibinfo  {journal} {Phys. Rev. D}\ }\textbf {\bibinfo {volume} {48}},\
  \bibinfo {pages} {3743} (\bibinfo {year} {1993})},\ \Eprint
  {http://arxiv.org/abs/hep-th/9306069} {arXiv:hep-th/9306069 [hep-th]}
  \BibitemShut {NoStop}%
\bibitem [{\citenamefont {{Smarr}}(1973)}]{Smarr73}%
  \BibitemOpen
  \bibfield  {author} {\bibinfo {author} {\bibfnamefont {L.}~\bibnamefont
  {{Smarr}}},\ }\href {\doibase 10.1103/PhysRevLett.30.71} {\bibfield
  {journal} {\bibinfo  {journal} {Phys. Rev. Lett.}\ }\textbf {\bibinfo
  {volume} {30}},\ \bibinfo {pages} {71} (\bibinfo {year} {1973})}\BibitemShut
  {NoStop}%
\bibitem [{\citenamefont {{Bardeen}}(1973)}]{Bardeen73}%
  \BibitemOpen
  \bibfield  {author} {\bibinfo {author} {\bibfnamefont {J.~M.}\ \bibnamefont
  {{Bardeen}}},\ }in\ \href@noop {} {\emph {\bibinfo {booktitle} {Black Holes
  (Les Astres Occlus)}}},\ \bibinfo {editor} {edited by\ \bibinfo {editor}
  {\bibfnamefont {C.}~\bibnamefont {{Dewitt}}}\ and\ \bibinfo {editor}
  {\bibfnamefont {B.~S.}\ \bibnamefont {{Dewitt}}}}\ (\bibinfo {year} {1973})\
  pp.\ \bibinfo {pages} {215--239}\BibitemShut {NoStop}%
\bibitem [{\citenamefont {{Komar}}(1959)}]{Komar1959}%
  \BibitemOpen
  \bibfield  {author} {\bibinfo {author} {\bibfnamefont {A.}~\bibnamefont
  {{Komar}}},\ }\href {\doibase 10.1103/PhysRev.113.934} {\bibfield  {journal}
  {\bibinfo  {journal} {Phys. Rev.}\ }\textbf {\bibinfo {volume} {113}},\
  \bibinfo {pages} {934} (\bibinfo {year} {1959})}\BibitemShut {NoStop}%
\bibitem [{\citenamefont {{Dadhich}}(2020)}]{Dadhich20:JCAP}%
  \BibitemOpen
  \bibfield  {author} {\bibinfo {author} {\bibfnamefont {N.}~\bibnamefont
  {{Dadhich}}},\ }\href {\doibase 10.1088/1475-7516/2020/04/035} {\bibfield
  {journal} {\bibinfo  {journal} {JCAP}\ }\textbf {\bibinfo {volume} {2020}},\
  \bibinfo {eid} {035} (\bibinfo {year} {2020})},\ \Eprint
  {http://arxiv.org/abs/1903.03436} {arXiv:1903.03436 [gr-qc]} \BibitemShut
  {NoStop}%
\bibitem [{\citenamefont {{Dadhich}}\ \emph {et~al.}(2023)\citenamefont
  {{Dadhich}}, \citenamefont {{Goswami}},\ and\ \citenamefont
  {{Hansraj}}}]{dago}%
  \BibitemOpen
  \bibfield  {author} {\bibinfo {author} {\bibfnamefont {N.}~\bibnamefont
  {{Dadhich}}}, \bibinfo {author} {\bibfnamefont {R.}~\bibnamefont
  {{Goswami}}}, \ and\ \bibinfo {author} {\bibfnamefont {C.}~\bibnamefont
  {{Hansraj}}},\ }\href {\doibase 10.48550/arXiv.2304.10197} {\bibfield
  {journal} {\bibinfo  {journal} {arXiv e-prints}\ ,\ \bibinfo {eid}
  {arXiv:2304.10197}} (\bibinfo {year} {2023})},\ \Eprint
  {http://arxiv.org/abs/2304.10197} {arXiv:2304.10197 [gr-qc]} \BibitemShut
  {NoStop}%
\bibitem [{\citenamefont {{Wagh}}\ \emph {et~al.}(1985)\citenamefont {{Wagh}},
  \citenamefont {{Dhurandhar}},\ and\ \citenamefont {{Dadhich}}}]{Wagh85}%
  \BibitemOpen
  \bibfield  {author} {\bibinfo {author} {\bibfnamefont {S.~M.}\ \bibnamefont
  {{Wagh}}}, \bibinfo {author} {\bibfnamefont {S.~V.}\ \bibnamefont
  {{Dhurandhar}}}, \ and\ \bibinfo {author} {\bibfnamefont {N.}~\bibnamefont
  {{Dadhich}}},\ }\href {\doibase 10.1086/162952} {\bibfield  {journal}
  {\bibinfo  {journal} {Astrophys J.}\ }\textbf {\bibinfo {volume} {290}},\
  \bibinfo {pages} {12} (\bibinfo {year} {1985})}\BibitemShut {NoStop}%
\bibitem [{\citenamefont {{Dadhich}}\ \emph {et~al.}(2018)\citenamefont
  {{Dadhich}}, \citenamefont {{Tursunov}}, \citenamefont {{Ahmedov}},\ and\
  \citenamefont {{Stuchl{\'\i}k}}}]{Dadhich18}%
  \BibitemOpen
  \bibfield  {author} {\bibinfo {author} {\bibfnamefont {N.}~\bibnamefont
  {{Dadhich}}}, \bibinfo {author} {\bibfnamefont {A.}~\bibnamefont
  {{Tursunov}}}, \bibinfo {author} {\bibfnamefont {B.}~\bibnamefont
  {{Ahmedov}}}, \ and\ \bibinfo {author} {\bibfnamefont {Z.}~\bibnamefont
  {{Stuchl{\'\i}k}}},\ }\href {\doibase 10.1093/mnrasl/sly073} {\bibfield
  {journal} {\bibinfo  {journal} {Mon. Not. Roy. Astron. Soc.}\ }\textbf
  {\bibinfo {volume} {478}},\ \bibinfo {pages} {L89} (\bibinfo {year}
  {2018})},\ \Eprint {http://arxiv.org/abs/1804.09679} {arXiv:1804.09679
  [astro-ph.HE]} \BibitemShut {NoStop}%
\bibitem [{\citenamefont {{Shaymatov}}\ \emph {et~al.}(2022)\citenamefont
  {{Shaymatov}}, \citenamefont {{Sheoran}}, \citenamefont {{Becerril}},
  \citenamefont {{Nucamendi}},\ and\ \citenamefont {{Ahmedov}}}]{Shaymatov22b}%
  \BibitemOpen
  \bibfield  {author} {\bibinfo {author} {\bibfnamefont {S.}~\bibnamefont
  {{Shaymatov}}}, \bibinfo {author} {\bibfnamefont {P.}~\bibnamefont
  {{Sheoran}}}, \bibinfo {author} {\bibfnamefont {R.}~\bibnamefont
  {{Becerril}}}, \bibinfo {author} {\bibfnamefont {U.}~\bibnamefont
  {{Nucamendi}}}, \ and\ \bibinfo {author} {\bibfnamefont {B.}~\bibnamefont
  {{Ahmedov}}},\ }\href {\doibase 10.1103/PhysRevD.106.024039} {\bibfield
  {journal} {\bibinfo  {journal} {Phys. Rev. D}\ }\textbf {\bibinfo {volume}
  {106}},\ \bibinfo {eid} {024039} (\bibinfo {year} {2022})}\BibitemShut
  {NoStop}%
\bibitem [{\citenamefont {{Richartz}}\ and\ \citenamefont
  {{Saa}}(2013)}]{Richartz13}%
  \BibitemOpen
  \bibfield  {author} {\bibinfo {author} {\bibfnamefont {M.}~\bibnamefont
  {{Richartz}}}\ and\ \bibinfo {author} {\bibfnamefont {A.}~\bibnamefont
  {{Saa}}},\ }\href {\doibase 10.1103/PhysRevD.88.044008} {\bibfield  {journal}
  {\bibinfo  {journal} {Phys. Rev. D}\ }\textbf {\bibinfo {volume} {88}},\
  \bibinfo {eid} {044008} (\bibinfo {year} {2013})},\ \Eprint
  {http://arxiv.org/abs/1306.3137} {arXiv:1306.3137 [gr-qc]} \BibitemShut
  {NoStop}%
\bibitem [{\citenamefont {{Cardoso}}\ \emph {et~al.}(2015)\citenamefont
  {{Cardoso}}, \citenamefont {{Brito}},\ and\ \citenamefont
  {{Rosa}}}]{Cardoso15}%
  \BibitemOpen
  \bibfield  {author} {\bibinfo {author} {\bibfnamefont {V.}~\bibnamefont
  {{Cardoso}}}, \bibinfo {author} {\bibfnamefont {R.}~\bibnamefont {{Brito}}},
  \ and\ \bibinfo {author} {\bibfnamefont {J.~L.}\ \bibnamefont {{Rosa}}},\
  }\href {\doibase 10.1103/PhysRevD.91.124026} {\bibfield  {journal} {\bibinfo
  {journal} {Phys. Rev. D}\ }\textbf {\bibinfo {volume} {91}},\ \bibinfo {eid}
  {124026} (\bibinfo {year} {2015})},\ \Eprint
  {http://arxiv.org/abs/1505.05509} {arXiv:1505.05509 [gr-qc]} \BibitemShut
  {NoStop}%
\bibitem [{\citenamefont {{Cardoso}}\ \emph {et~al.}(2017)\citenamefont
  {{Cardoso}}, \citenamefont {{Pani}},\ and\ \citenamefont {{Yu}}}]{Cardoso17}%
  \BibitemOpen
  \bibfield  {author} {\bibinfo {author} {\bibfnamefont {V.}~\bibnamefont
  {{Cardoso}}}, \bibinfo {author} {\bibfnamefont {P.}~\bibnamefont {{Pani}}}, \
  and\ \bibinfo {author} {\bibfnamefont {T.-T.}\ \bibnamefont {{Yu}}},\ }\href
  {\doibase 10.1103/PhysRevD.95.124056} {\bibfield  {journal} {\bibinfo
  {journal} {Phys. Rev. D}\ }\textbf {\bibinfo {volume} {95}},\ \bibinfo {eid}
  {124056} (\bibinfo {year} {2017})},\ \Eprint
  {http://arxiv.org/abs/1704.06151} {arXiv:1704.06151 [gr-qc]} \BibitemShut
  {NoStop}%
\bibitem [{\citenamefont {{Day}}\ and\ \citenamefont
  {{McDonald}}(2019)}]{Day19JCAP}%
  \BibitemOpen
  \bibfield  {author} {\bibinfo {author} {\bibfnamefont {F.~V.}\ \bibnamefont
  {{Day}}}\ and\ \bibinfo {author} {\bibfnamefont {J.~I.}\ \bibnamefont
  {{McDonald}}},\ }\href {\doibase 10.1088/1475-7516/2019/10/051} {\bibfield
  {journal} {\bibinfo  {journal} {JCAP}\ }\textbf {\bibinfo {volume} {2019}},\
  \bibinfo {eid} {051} (\bibinfo {year} {2019})},\ \Eprint
  {http://arxiv.org/abs/1904.08341} {arXiv:1904.08341 [hep-ph]} \BibitemShut
  {NoStop}%
\bibitem [{\citenamefont {{Chadha-Day}}\ \emph {et~al.}(2022)\citenamefont
  {{Chadha-Day}}, \citenamefont {{Garbrecht}},\ and\ \citenamefont
  {{McDonald}}}]{Chadha-Day22JCAP}%
  \BibitemOpen
  \bibfield  {author} {\bibinfo {author} {\bibfnamefont {F.}~\bibnamefont
  {{Chadha-Day}}}, \bibinfo {author} {\bibfnamefont {B.}~\bibnamefont
  {{Garbrecht}}}, \ and\ \bibinfo {author} {\bibfnamefont {J.~I.}\ \bibnamefont
  {{McDonald}}},\ }\href {\doibase 10.1088/1475-7516/2022/12/008} {\bibfield
  {journal} {\bibinfo  {journal} {JCAP}\ }\textbf {\bibinfo {volume} {2022}},\
  \bibinfo {eid} {008} (\bibinfo {year} {2022})},\ \Eprint
  {http://arxiv.org/abs/2207.07662} {arXiv:2207.07662 [hep-ph]} \BibitemShut
  {NoStop}%
\end{thebibliography}%

 \end{document}